\documentclass[twocolumn,aps,prb,amsmath,amssymb,nobibnotes]{revtex4}

\usepackage[]{graphicx}
\usepackage{bm}

\newcommand{\DDt}[1]{\mathrm{D}_t{#1}}

\newcommand{\ddt}[1]{\frac{\mathrm{d}{#1}}{\mathrm{d}t}}
\newcommand{\pp}[2]{\partial_{#1}{#2}}
\newcommand{\ppt}[1]{\pp{t}{#1}}
\renewcommand{\vec}[1]{\ensuremath\boldsymbol{\mathrm{#1}}}
\newcommand{\uvec}[1]{\vec{\hat{#1}}}

\newcommand{\nablab}{\boldsymbol{\nabla}}
\newcommand{\cdotb}{\boldsymbol{\cdot}}

\newcommand{\timesb}{\boldsymbol{\times}}

\newcommand{\spar}{\shortparallel}
\newcommand{\abs}[1]{\lvert{#1}\rvert}
\newcommand{\ensav}[1]{\left\langle{#1}\right\rangle}
\newcommand{\mean}[1]{\overline{#1}}
\newcommand{\intd}[2]{\int{#2}\,\mathrm{d}{#1}}
\newcommand{\order}[1]{\mathcal{O}({#1})}

\newcommand{\Figref}[1]{Fig.~\ref{#1}}

\newcommand{\Tabref}[1]{Table~\ref{#1}}

\newcommand{\Eqref}[1]{Eq.~\eqref{#1}}
\newcommand{\Eqsref}[1]{Eqs.~\eqref{#1}}
\newcommand{\Ref}[1]{Ref.~\onlinecite{#1}}

\newcommand{\Secref}[1]{Sec.~\ref{#1}}
\newcommand{\Secsref}[1]{Secs.~\ref{#1}}

\begin{document}

\title{Anomalous diffusion, clustering, and pinch of impurities in plasma edge turbulence}

\author{M.~Priego}
\altaffiliation[Also at ]{Department of Mathematics, Technical University of Denmark and 
Departamento de Fluidos y Calor, Universidad Pontificia Comillas--ICAI}
\email[E-mail address: ]{martin.priego@telefonica.net}
\author{O.~E.~Garcia}
\author{V.~Naulin}
\author{J.~Juul~Rasmussen}
\affiliation{Association EURATOM--Ris{\o} National Laboratory,
OPL-128 Ris{\o}, DK-4000 Roskilde, Denmark}

\date{January 7, 2005}

\begin{abstract}
The turbulent transport of impurity particles in plasma edge turbulence is
investigated. The impurities are modeled as a passive fluid advected by the
electric and polarization drifts, while the ambient plasma turbulence is modeled
using the two-dimensional Hasegawa--Wakatani paradigm for resistive drift-wave
turbulence. The features of the turbulent transport of impurities are
investigated by numerical simulations using a novel code that applies
semi-Lagrangian pseudospectral schemes. The diffusive character of the turbulent
transport of ideal impurities is demonstrated by relative-diffusion analysis of
the evolution of impurity puffs. Additional effects appear for inertial
impurities as a consequence of compressibility. First, the density of inertial
impurities is found to correlate with the vorticity of the electric drift
velocity, that is, impurities cluster in vortices of a precise orientation
determined by the charge of the impurity particles. Second, a radial pinch
scaling linearly with the mass--charge ratio of the impurities is
discovered. Theoretical explanation for these observations is obtained by
analysis of the model equations.
\end{abstract}

\maketitle

\section{Introduction}\label{sec:intro}
One of the persistent problems in the development of a magnetic fusion reactor
is the degradation of plasma confinement caused by turbulent transport. In spite
of the strong confining magnetic field, high levels of energy and
charged-particle transport across magnetic field lines are observed in a variety
of experimental configurations, such as tokamaks and stellarators. This
transport has received the adjective ``anomalous,'' as its features cannot be
explained by the classical or neoclassical collisional theories. Instead,
anomalous transport is generally attributed to small-scale, electrostatic plasma
turbulence.\cite{wagn93,hort99} In particular, the transport in the plasma edge
is believed to be largely controlled by drift-wave turbulence, which appears
naturally in this region as a result of the strong radial pressure gradient. The
drift-wave transport mechanism has been the subject of much theoretical and
experimental research over the past several decades, and was recently reviewed
by Horton.\cite{hort99} Nevertheless, much of this transport
phenomenon is yet to be fully understood.

Another important problem in the quest for controlled thermonuclear fusion that
is linked to turbulence concerns the impurity content of the plasma. The
properties of fusion plasmas are strongly influenced by the presence of
impurities. For instance, impurities enhance radiative energy losses and dilute
the hydrogen fuel within the hot plasma center, at the same time being the main
agents in the erosion and deposition processes that affect plasma-facing
components.\cite{iter99} Consequently, the understanding of the transport
mechanisms that control the impurity content in the plasma is of crucial
importance for the performance and safe operation of future fusion devices.

The transport of impurity particles across the magnetic field is related to that
of the bulk plasma, since the motion perpendicular to the magnetic field of all
charged particles in the plasma is dominated by one same velocity field, the
electric drift velocity. In this sense, impurities and bulk plasma are partly
subject to a common mechanism of turbulent transport, which relates the problem
of confinement to that of controlling the impurity content. However, the
distinctive nature of impurity particles, notably their mass and electric
charge, as well as their different influence on the plasma allow for disparities
between the transport of impurities and that of the bulk plasma. In fact, recent
experiments not only reveal disparities in the transport of impurities and bulk
plasma,\cite{zast99} but also in that of different impurity species.\cite{dux03}
In such experimental investigations impurity transport is usually characterized
in terms of effective diffusion coefficients and drift velocities, the latter
also known as pinch velocities.\cite{wagn93,garb04} These transport coefficients
are typically determined by analysis of the temporal evolution of injected
impurity puffs. While the explanation for the diffusive character of turbulent
transport goes back to the work of Taylor and Richardson in the
1920s,\cite{tayl21,rich26} the origin of the anomalous pinch velocities is still
a matter of intense research.\cite{garb04} Moreover, the inward pinch observed
in laboratory experiments tends to concentrate the impurities in the center of
the plasma column,\cite{wagn93} precisely where they are least wanted, and thus
constitutes one of the major unresolved transport problems in fusion plasma
physics.

In this contribution we investigate the turbulent transport of impurities in a
generic model for plasma edge turbulence. Specifically, the investigations are
based on the simple and extensively studied model for resistive drift-wave
turbulence proposed by Hasegawa and Wakatani.\cite{hase83} For the impurities, we
use a consistent passive-fluid model that takes into account impurity-particle
inertia, that is to say, it contemplates the polarization drift. This enables
the model to exhibit different transport features depending on the mass and
charge of the impurities, as observed in laboratory experiments. In particular,
the polarization drift brings compressibility into the model, giving rise to
subtle yet qualitatively important effects. The investigations are carried out
by numerical simulations of the model equations using a newly developed code that
applies semi-Lagrangian pseudospectral schemes. We proceed separately for ideal
(i.e., massless) impurities and inertial ones. In the case of ideal impurities,
we focus on the phenomenon of turbulent diffusion, which is studied from the
relative-diffusion perspective. For inertial impurities, we concentrate on the
new effects that arise from compressibility. First, the density of inertial
impurities is found to correlate with the vorticity of the electric drift
velocity, that is, impurities cluster in vortices of a precise orientation
determined by the charge of the impurity particles. This behavior is a direct
consequence of compression by the polarization drift and turbulent
mixing. Second, we discover a radial pinch that scales linearly with the
mass--charge ratio of the impurities. An understanding of this phenomenon is
obtained by analysis of the evolution equation for the global impurity flux.

This paper is organized as follows. In \Secref{sec:model} the model for
resistive drift-wave turbulence and the passive-fluid model for the impurities
are described. The diffusive effect of turbulence is demonstrated in
\Secref{sec:ideal} for the case of ideal impurities. The effects that arise from
impurity inertia, the clustering and the anomalous radial pinch, are
respectively presented in \Secsref{sec:cluster} and \ref{sec:pinch}. The results
are summarized and discussed in \Secref{sec:conclus}. Lastly, the Appendix is
devoted to a brief description of the numerical schemes applied in the
simulations.

\section{Model equations}\label{sec:model}
In this section we provide a brief presentation of the resistive drift-wave
paradigm used for modeling the turbulence at the plasma edge. We also describe
the passive-fluid model for the impurity species, which takes into account
impurity-particle inertia.

\subsection{Resistive drift-wave turbulence}\label{ssec:model_dwt}
As a basic model for plasma edge turbulence we consider the two-dimensional (2D)
Hasegawa--Wakatani (HW) model for resistive drift-wave turbulence:\cite{hase83}
\begin{subequations}\label{eq:hw}
\begin{align}
\label{eq:hw_n} \DDt{(n-x)}&  = \mathcal{C} (\varphi-n) + \mu_n \nabla_\perp^2 n,\\[1ex]
\label{eq:hw_w} \DDt{\omega}& = \mathcal{C} (\varphi-n) + \mu_\omega \nabla_\perp^2 \omega,
\end{align}
where 
\begin{align}
\label{eq:wdef} \omega &\equiv\nabla_\perp^2 \varphi,\\[1ex]
\label{eq:ddt} \DDt{} &\equiv\ppt{}+\uvec{z}\timesb\nablab_\perp\varphi \cdotb\nablab_\perp.
\end{align}
\end{subequations}
Here, $n$ denotes the fluctuating component of the plasma density, $\varphi$ is
the electrostatic potential, and $\omega$ is the vorticity of the electric drift
velocity $\vec{v}_E=\uvec{z}\timesb\nablab_\perp\varphi$.

The HW model assumes an unperturbed, uniform magnetic field in the
$z$-direction, that is, $\vec{B}=B_0\uvec{z}$. The $x$- and $y$-directions
correspond respectively to the radial and the poloidal directions in a slab at
the edge of the confined plasma. The electrons are assumed isothermal with
temperature $T_e$, whereas the ions are considered cold. The equilibrium plasma
density $n_0$ is assumed to be homogeneous in the poloidal direction and to
decrease in the radial direction in such a way that the equilibrium density
length scale $L_n\equiv n_0/\abs{\mathrm{d}n_0/\mathrm{d}x}$ is constant.
Hence, in this thin-layer approximation the second term on the left-hand side of
\Eqref{eq:hw_n} actually represents the advection of the inhomogeneous
equilibrium plasma density by the electric drift.

The quantities in \Eqsref{eq:hw} have been made dimensionless using the 
gyro-Bohm normalization:
\begin{equation*}
\frac{x}{\rho_s}\to x,\ 
\frac{y}{\rho_s}\to y,\ 
t\,\frac{c_s}{L_n}\to t,\ 
\frac{n}{n_0}\frac{L_n}{\rho_s}\to n,\ 
\frac{e\varphi}{T_e}\frac{L_n}{\rho_s}\to \varphi.
\end{equation*}
Here, $c_s\equiv(T_e/m_i)^{1/2}$ is the sound speed and $\rho_s\equiv(m_i T_e)^{1/2}/e B_0$
is the the drift scale, that is, the ion gyroradius at the sound speed.
The parameter $\mathcal{C}$ coupling \Eqsref{eq:hw_n} and \eqref{eq:hw_w}
is known as the adiabaticity parameter, and is defined as
\begin{equation*}
\mathcal{C} \equiv \frac{T_e}{e^2 \eta_\spar n_0} \frac{k_\spar^2}{c_s/L_n}.
\end{equation*}
Here, $\eta_\spar$ is the parallel resistivity and $k_\spar$ is the single
parallel wave number that is considered in the reduction to a 2D model.  In the
strongly collisional limit $\mathcal{C}\to0$, \Eqref{eq:hw_w} reduces to the 2D
Navier--Stokes vorticity equation. In this so-called hydrodynamic limit the
equations become decoupled and the plasma density is passively advected by the
electric drift velocity. In the weakly collisional limit $\mathcal{C}\to\infty$,
the adiabatic electron response yields Boltzmann-distributed density
perturbations, $n\approx\varphi$, and the HW model reduces to the well-known
Hasegawa--Mima equation.\cite{hase78}

The diffusive terms in \Eqsref{eq:hw_n} and \eqref{eq:hw_w} correspond
respectively to the collisional diffusion of electrons and the ion viscosity.
These dissipative processes are essential for the stabilization of resistive
drift modes as well as for the feasibility of numerically simulating the HW
model. Linear analysis of the HW model reveals the existence of unstable modes
for $0<\mathcal{C}<\infty$, with all modes being unstable in the absence of
dissipation.\cite{garc01,cama95} It is the combination of this linear
instability with nonlinear interaction through the advective terms and
dissipation at small scales that leads to quasistationary turbulent states
within the HW model. In particular, when initialized with a broadband set of
small-amplitude, random-phase modes, the system evolves in two stages: first, a
linear phase in which the linearly unstable modes grow and the damped ones
decay; second, a phase of nonlinear saturation leading to a quasistationary
turbulent state whose statistics solely depend on the parameters of the HW
model.

Two relevant nonlinear invariants in the HW model are the total energy $E$ and
the generalized enstrophy $U$, defined as
\begin{align*}
E &\equiv \frac{1}{2}\intd{\vec{x}}{(n^2 + \abs{\nablab_\perp\varphi}^2)},\\[1ex]
U &\equiv \frac{1}{2}\intd{\vec{x}}{(n - \omega)^2}.
\end{align*}
The evolution of these quantities within the HW model is
governed by
\begin{align*}
\ddt{E}& = \Gamma_n - \Gamma_c - \mathcal{D}^E,\\[1ex]
\ddt{U}& = \Gamma_n - \mathcal{D}^U,
\end{align*}
where
\begin{align*}
\Gamma_n& \equiv -\intd{\vec{x}}{n\pp{y}{\varphi}},\\[0.5ex]
\Gamma_c& \equiv \mathcal{C} \intd{\vec{x}}{(n-\varphi)^2},\\[0.5ex]
\mathcal{D}^E& \equiv \intd{\vec{x}}{(\mu_n\abs{\nablab_\perp n}^2+
\mu_\omega\abs{\nablab_\perp \omega}^2)},\\[0.5ex]
\mathcal{D}^U& \equiv \intd{\vec{x}}{\nablab_\perp(n-\omega)\cdotb
\nablab_\perp(\mu_n n-\mu_\omega\omega)}.
\end{align*}
Hence, $\Gamma_n$ is the only source in the system, and corresponds to energy
extracted from the equilibrium density gradient. Conversely, $\Gamma_c$ is a
sink corresponding to resistive dissipation through the parallel current, while
$\mathcal{D}^E$ and $\mathcal{D}^U$ are sinks arising from collisionality and
viscosity. Being the sole source of energy and generalized enstrophy source
within the HW model, $\Gamma_n$ is in practice positive definite.\cite{cama95}
This fact is used in \Secref{sec:pinch} to identify a source for radial drift of
impurities.

\subsection{Impurity passive-fluid model}\label{ssec:model_imp}
Impurities are here regarded as a passive species, that is, they are advected by
the turbulence but do not have any influence on it. This is a reasonable
approximation for the situation in which the density of impurity particles is
much lower than that of the bulk plasma. The modeling of the dynamics of
impurities is consistent with that of the turbulence, in that it also follows
from the drift-ordered fluid approach and the assumption of unperturbed, uniform
magnetic field.

The impurities are characterized by a density $\theta$ and a flow velocity
$\vec{v}_\theta$. Just as in the case of the ions in the bulk plasma, the
impurities are assumed cold, with their motion parallel to the magnetic field
likewise being neglected. Therefore, the motion of the impurity fluid comprises
only the electric drift $\vec{v}_E$ and the polarization drift $\vec{v}_p$.
Under the same normalization applied to the turbulence model, the flow velocity
of the impurities is hence given by
\begin{equation*}
\vec{v}_\theta = \vec{v}_E + \vec{v}_p = 
\vec{\hat z}\timesb\nablab_\perp\varphi - \zeta\DDt{\nablab_\perp \varphi},
\end{equation*}
where 
\begin{equation*}
\zeta\equiv\frac{m_\theta}{q_\theta}\frac{e}{m_i}\frac{\rho_s}{L_n}.
\end{equation*}
Here, $m_\theta$ and $q_\theta$ are respectively the mass and the electric
charge of the impurity particles. The impurity model resulting from the
continuity equation is thus given by
\begin{equation}\label{eq:impur}
\DDt{\theta} -
\nablab_\perp\cdotb\bigl(\theta\zeta\DDt{\nablab_\perp\varphi}\bigr) =
\mu_\theta\nabla_\perp^2\theta,
\end{equation}
where the term on the right-hand side arises from collisional diffusion.

The polarization drift, which represents the effect of impurity-particle
inertia, constitutes a higher-order correction to the electric drift
velocity. The influence of the former on the dynamics of impurities is
determined by the dimensionless parameter $\zeta$, which varies with the
mass--charge ratio of the impurity particles relative to that of the bulk-plasma
ions. For ideal impurities $\zeta=0$ and the impurities are solely advected by
the electric drift. For inertial impurities $\zeta$ has a nonzero value and the
polarization drift can play a role in the dynamics. The value of $\zeta$ is in
any case small, since the length scale quotient $\rho_s/L_n$ is by
assumption small. Nevertheless, the polarization drift adds important
qualitative features to the dynamics of impurities. In the case of static, 
uniform magnetic field the electric drift is incompressible, whereas the
polarization drift is compressible. Hence, the inclusion of the polarization
drift in the impurity model allows for genuinely compressible effects, such as
those described in \Secsref{sec:cluster} and \ref{sec:pinch}.

\section{Diffusion of ideal impurities}\label{sec:ideal}
Theoretical studies of the diffusive effect of plasma edge turbulence on
passively advected impurities have previously relied on analysis of the absolute
dispersion of test particles.\cite{basu03a,basu03b} The present approach is
necessarily different, as the impurities are not modeled here as particles but
as a fluid. In this section we investigate the turbulent diffusion of ideal
impurities by relative-diffusion analysis of the evolution of impurity
puffs.\cite{csan73,mikk87}

\subsection{Relative diffusion}
We consider the release of a puff of ideal impurities into the turbulent field,
which is a saturated turbulent state described by the HW model. The amount of
impurities released is
\begin{equation*}
Q\equiv\intd{\vec{x}}{\theta},
\end{equation*}
and the position $\vec{c}$ of the center of mass of the impurity puff at any time instant
is given by
\begin{equation*}
\vec{c}\equiv \frac{1}{Q} \intd{\vec{x}}{\vec{x}\theta}.
\end{equation*}
We define the relative position vector $\vec{x}'\equiv\vec{x}-\vec{c}$, which
indicates position in a reference frame moving with the center of mass of the
puff. The average evolution of impurity puffs can then be characterized by the
following ensemble-averaged dispersion tensors:
\begin{align*}
\Sigma_{ij}&\equiv\frac{1}{Q}\ensav{\intd{\vec{x}}{x_i x_j \theta}},\\[1ex]
S_{ij}&\equiv\frac{1}{Q}\ensav{\intd{\vec{x}}{x'_i x'_j \theta}},\\[1ex]
M_{ij}&\equiv\ensav{c_i c_j}.
\end{align*}
Here, the components of vectors and tensors have been indicated by subindices
and the angular brackets stand for the ensemble average, that is, the average
over many releases. The absolute-dispersion tensor $\Sigma_{ij}$ measures the
dispersion with respect to the origin of the fixed frame, whereas the
relative-dispersion tensor $S_{ij}$ measures the dispersion relative to the
center of mass. The tensor $M_{ij}$ quantifies the ``meandering'' of the center
of mass. The dispersion tensors are related by
\begin{equation*}
\Sigma_{ij} = S_{ij} + M_{ij},
\end{equation*}
which means that absolute dispersion consists of relative dispersion and
meandering of the center of mass. Relative-diffusion analysis focuses on the
evolution of $S_{ij}$, that is, on the spreading of the puff relative to its
center of mass. Specifically, it is usually the diagonal elements $S_{ii}$ that
are of interest once the reference axes are adequately selected, in our case
those corresponding to the radial and poloidal directions. In view of
the above relation, the meandering of the center of mass is eliminated in
relative-diffusion analysis, which thus leads to better statistics in comparison
with those of absolute diffusion.

The evolution of the dispersion tensors depends on the initial size and shape of
the puffs, the statistical properties of the turbulent velocity field, and the
collisional diffusivity $\mu_\theta$. Under the assumption that $\mu_\theta$ is
small, the evolution of $S_{ii}$ for initially concentrated puffs follows three
stages: first, a phase dominated by collisional diffusion that lasts while the
puff is much smaller than the turbulent structures and is characterized by
\begin{equation*}
\ddt{S_{ii}} \approx 2\mu_\theta;
\end{equation*}
second, a transitional phase in which the size of the puff is comparable to that of
the turbulent structures and $S_{ii}$ typically evolve faster than linearly with
time (see \Ref{csan73} for further discussions); and last, a phase dominated
by turbulent diffusion that starts when the puff is large compared to the
turbulent structures and during which the evolution of $S_{ii}$ can be
characterized by effective diffusivities $D_i$ such that
\begin{equation}
\label{eq:eff_diff}\ddt{S_{ii}} = 2D_i.
\end{equation}
These diffusivities solely depend on the statistical properties of the turbulent
velocity field---more precisely, they are related to the mean squared velocities
and the Lagrangian relative time scale.

The last phase, dominated by turbulent diffusion, is most relevant for us, as it
eventually dominates the transport of impurities. In fact, the effective
diffusivities $D_i$ that characterize the phase of turbulent diffusion are
equivalent in their definition to those employed in the experimental modeling of
anomalous impurity transport.\cite{wagn93,garb04} Because in this last stage the 
diagonal elements $S_{ii}$ grow linearly according to \Eqref{eq:eff_diff}, the effective
diffusivities $D_i$ can easily be obtained from the asymptotic evolution of
$S_{ii}$. This is the procedure used here for estimating the effective
diffusivities of the turbulent field acting on ideal impurities. In principle,
any initial puffs could be used for the determination of the diffusivities. From
the discussion above, large initial puffs rapidly experience the final phase of
turbulent diffusion, and thus provide the effective diffusivities shortly after
their release. However, care must be taken in numerical simulations that the
puffs do not reach the boundaries of the simulation domain, as this would result
in underestimates for the diffusivities. In order to determine the effective
diffusivity in one direction, we take Gaussian stripes in the perpendicular
direction as initial puffs and use periodic boundary conditions. In this way,
the puffs are as large as possible in one direction while the conclusions of the
simulations remain valid. Moreover, because Gaussians are similarity solutions
to the diffusion equation, the evolution of $S_{ii}$ would be perfectly linear
if the effect of turbulence were purely diffusive from the start.

\subsection{Simulation results}
The HW model for plasma edge turbulence, \Eqsref{eq:hw}, and the model for ideal
impurities, \Eqref{eq:impur} with $\zeta=0$, were simultaneously solved
numerically using the schemes described in the Appendix. The simulations were
carried out on a doubly periodic square domain of side length $40$ using
$512\times512$ grid points. The viscosity $\mu_\omega$ and the collisional
diffusivities $\mu_n$ and $\mu_\theta$ were set to the common value $0.02$ in
all the simulations presented in this paper. We considered four different values
for the adiabaticity parameter $\mathcal{C}$ in the HW model: $0.5$, $1$, $2$,
and $4$. As noted in the preceding section, low values of $\mathcal{C}$
correspond to hydrodynamic-like regimes, whereas high values correspond to
nearly adiabatic regimes. In this respect, we regard $\mathcal{C}=1$ as a
transitional value, representative of an intermediate regime. An initial
turbulent state was achieved by letting a uniform distribution of
small-amplitude, random-phase modes evolve for a sufficiently long time in order
to reach a quasistationary saturated turbulent regime. Gaussian stripes having
standard deviation $2$ were used as initial impurity puffs. For each value of
the adiabaticity parameter and each direction, radial and poloidal, $20$
releases of such stripes were simulated, and the dispersion tensors were
subsequently averaged.

\begin{figure}
\includegraphics{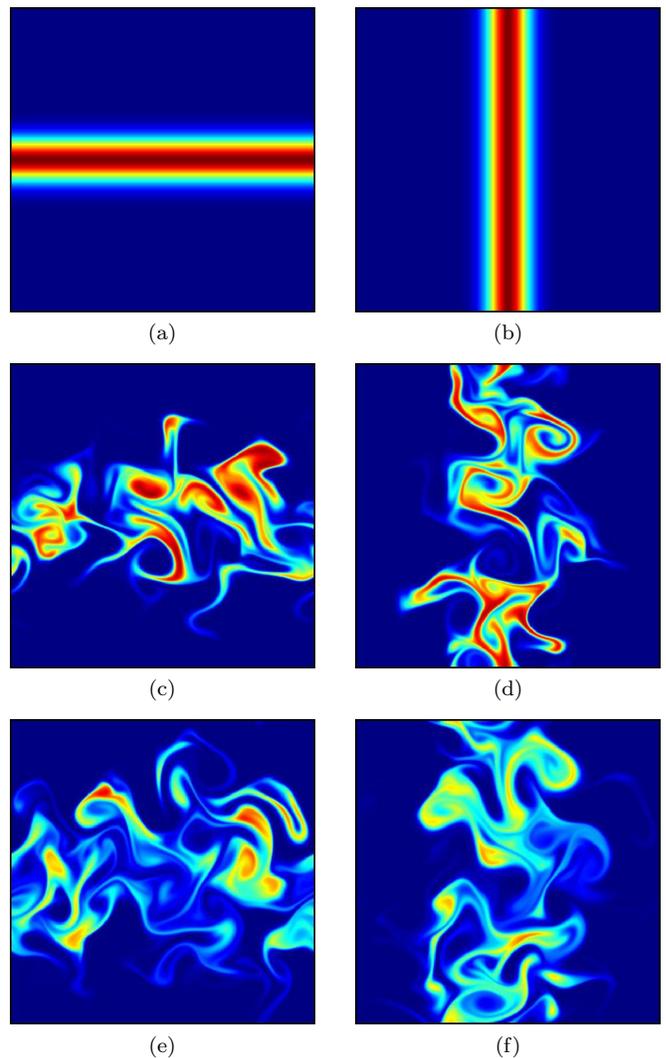}
\caption{\label{fig:1}%
(Color online) Evolution of Gaussian stripes of ideal impurities in a HW saturated 
turbulent state with $\mathcal{C}=1$: radial stripe at (a) $t=0$, (c) $t=6$, 
and (e) $t=12$; and poloidal stripe at (b) $t=0$, (d) $t=6$, and (f) $t=12$.}
\end{figure}

In \Figref{fig:1} we present snapshots of the evolution of two
perpendicular stripes of impurities released into the same saturated turbulent
field with $\mathcal{C}=1$. The anisotropy in the diffusion of impurities is
evident in the figure, as the poloidal spread of the puff in
\Figref{fig:1}(e) is significantly larger than the corresponding
radial spread in \Figref{fig:1}(f). Hence, diffusion acts in this case
faster in the poloidal direction than in the radial direction. The degree of
anisotropy is in fact controlled by the adiabaticity parameter, the turbulence
ranging from isotropic in hydrodynamic limit, $\mathcal{C}\to0$, to strongly
anisotropic in the adiabatic limit, $\mathcal{C}\to\infty$.\cite{basu03b}

\begin{figure}
\includegraphics{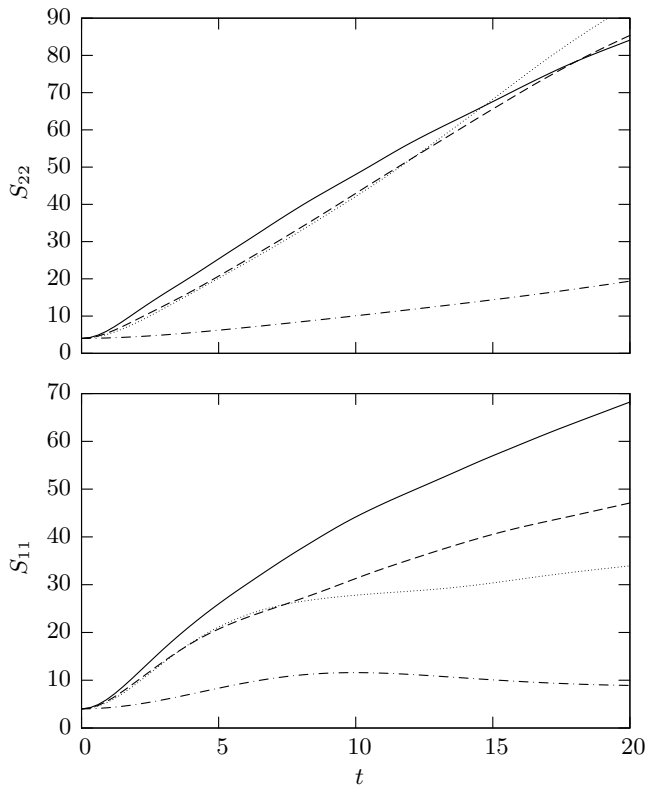}
\caption{\label{fig:2}%
Evolution of the relative radial dispersion $S_{11}$
and the relative poloidal dispersion $S_{22}$ of ideal impurity Gaussian stripes released 
in HW saturated turbulence with $\mathcal{C}=0.5$ (solid), $\mathcal{C}=1$
(dashed), $\mathcal{C}=2$ (dotted), and $\mathcal{C}=4$ (dash-dotted).}
\end{figure}

The evolution of the diagonal elements $S_{ii}$ of the relative dispersion
tensor for the four values of the adiabaticity parameter is shown in
\Figref{fig:2}. For the three smallest values of $\mathcal{C}$, it is possible
to distinguish the transitional and turbulent phases of growth in the plot for
relative radial dispersion. In the plot corresponding to poloidal dispersion
there is no clear distinction, as the evolution of $S_{22}$ very rapidly becomes
approximately linear. For the side length of our simulation domain, dispersion
can start becoming underestimated when $S_{ii}$ reach values of about
$65$. Nevertheless, for the three smallest values of $\mathcal{C}$ the linear
trend in $S_{ii}$ characterizing the phase of turbulent diffusion is established
before such a point is reached.  In contrast, the eventually linear regime is
neither apparent in the radial nor the poloidal dispersion plots for the highest
value of the adiabaticity parameter, $\mathcal{C}=4$. This different behavior is
attributed to the weaker turbulence at higher values of $\mathcal{C}$ and the
particular transport features of the adiabatic limit, the Hasegawa--Mima
equation.\cite{anni00}

Estimates for the effective diffusivities $D_i$ of the turbulence for the three
smallest values of $\mathcal{C}$ are obtained from linear fits to the linear
portions of the curves in \Figref{fig:2}. The estimated values for the
radial, $D_1$, and poloidal, $D_2$, effective diffusivities are presented in
\Tabref{tab:table1}. The values obtained here are qualitatively similar to
those calculated in \Ref{basu03b} by means of a test-particle,
absolute-diffusion approach. In particular, the decrease of $D_1$ with
$\mathcal{C}$ as well as the rise of $D_2$ with $\mathcal{C}$ for
$\mathcal{C}\ge1$ are here reproduced. We note that the accuracy of the present
approach could be improved by simulating the evolution of the impurity puffs for
longer times on a larger domain. The number of puff releases could as well be
increased in order to ensure convergence to the ensemble averages. We also note
that different values for $\mu_\omega$ and $\mu_n$ were used in the latter
reference, with collisional diffusion of impurities being discarded because of
their test-particle modeling.
\begin{table}
\caption{\label{tab:table1}%
Effective diffusivities in the radial direction $D_1$ and the poloidal
direction $D_2$ of HW saturated turbulence for different values of $\mathcal{C}$.}
\begin{ruledtabular}
\begin{tabular}{ccc}
$\mathcal{C}$ & $D_1$ & $D_2$\\
\hline
0.5 & 1.12 & 2.32\\
1 & 0.65 & 2.26\\
2 & 0.36 & 2.63
\end{tabular}
\end{ruledtabular}
\end{table}

\section{Clustering of inertial impurities}\label{sec:cluster}
The importance of inertia in the advection of particles by turbulent flows is
well known in fluid dynamics.\cite{prov99} As in magnetized plasmas, the
transport of inertial particles by fluids shows compressible features even when
the advecting flow is incompressible. Particle inertia has been shown to result
in clustering in vortical structures, and its role in phenomena such as
planetary formation has been suggested.\cite{brac99} While the dynamics of
impurity particles in magnetized plasmas are not governed by the same forces as
in fluids, inertia enters the dynamics of impurities in magnetized plasmas in a
similar way as the Coriolis force enters those of heavy impurities in rotating
fluids. Hence, corresponding clustering effects for inertial impurities in
magnetized plasmas should be expected. In this section we discover and explain
the correlation between impurity density and vorticity that results from the
compressibility introduced by the polarization drift in the case of inertial
impurities. In order to most clearly reveal this effect we consider a
homogeneous initial distribution of impurities, since such a distribution would
remain unchanged in the incompressible, that is, ideal-impurity case.

\subsection{Simulation results}
Simulations of the HW model for plasma edge turbulence, \Eqsref{eq:hw}, and the
impurity model, \Eqref{eq:impur}, were again performed on a doubly periodic
square domain of side length $40$ using $512\times512$ grid points. The adiabaticity
parameter $\mathcal{C}$ in the HW model was set to the transitional value $1$,
while we contemplated various values for the parameter 
$\zeta$, ranging from $-0.01$ to $0.05$. The impurity density field $\theta$ 
was initialized with the constant value $\theta_0=1$, and the initial turbulent 
fields were taken from a HW saturated turbulent state.

\begin{figure}
\includegraphics{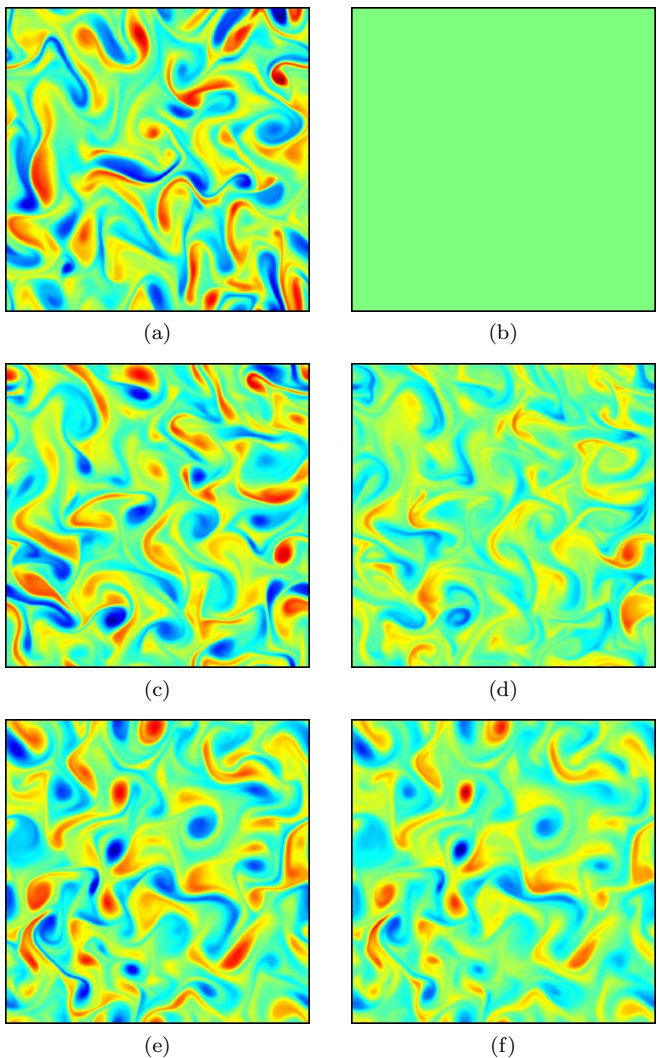}
\caption{\label{fig:3}%
(Color online) Evolution of the vorticity $\omega$ and the density $\theta$ of inertial 
impurities ($\zeta=0.01$) in a HW saturated turbulent state with $\mathcal{C}=1$: 
(a) $\omega$ and (b) $\theta$ at $t=0$,
(c) $\omega$ and (d) $\theta$ at $t=5$, and
(e) $\omega$ and (f) $\theta$ at $t=50$. Green corresponds to mean values, 
while red and blue represent positive and negative fluctuations respectively.}
\end{figure}

\begin{figure}
\includegraphics{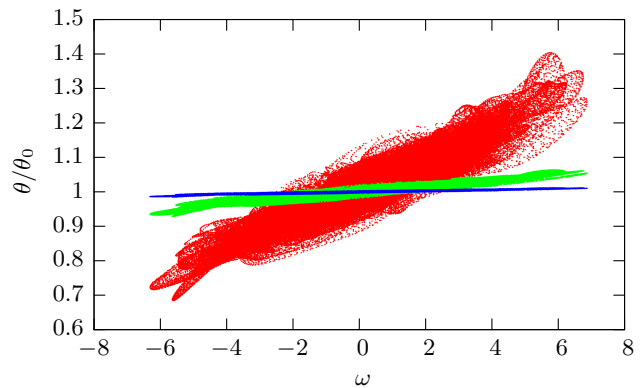}
\caption{\label{fig:4}%
(Color online) Scatter plot of vorticity $\omega$ and relative density 
$\theta/\theta_0$ of inertial impurities in a HW saturated turbulent state 
with $\mathcal{C}=1$ for $\zeta=0.05$ (red/steepest), 
$\zeta=0.01$ (green), and $\zeta=0.002$ (blue/flattest).}
\end{figure}

\begin{figure}
\includegraphics{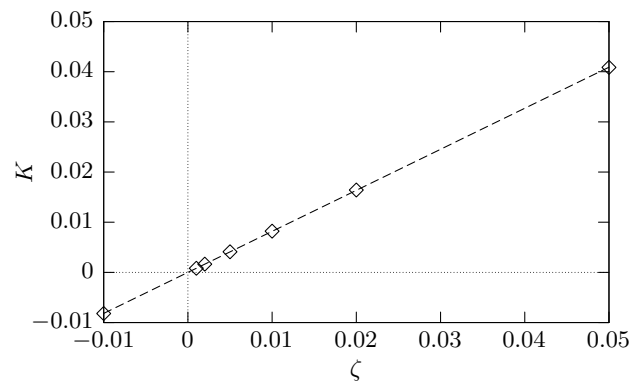}
\caption{\label{fig:5}%
Least-squares estimates of the coefficient $K$ as a function of $\zeta$ for 
the impurity-density--vorticity linear regression 
$\theta/\theta_0\approx1+K\omega$ in a HW saturated turbulent state 
with $\mathcal{C}=1$. The fitted line (dashed) corresponds to $K=0.82\zeta$.}
\end{figure}

In \Figref{fig:3} we present side by side snapshots of the evolution of
the vorticity $\omega$ of the electric drift velocity and the impurity density
field for the case $\zeta=0.01$.  Merely $5$ time units after the simulation is
started, there is already indication of a correlation between vorticity and
impurity density. At $t=50$, there is visually little difference between the two
fields. Figure~\ref{fig:4} shows a scatter plot of vorticity and relative
impurity density $\theta/\theta_0$ at $t=100$ for three different values of $\zeta$. 
The plot clearly suggests an approximate linear relation given by
\begin{equation*}
\theta/\theta_0\approx1+K\omega,
\end{equation*}
with the regression coefficient $K$ depending on $\zeta$. The least-squares
estimates of the coefficient $K$ are plotted as a function of the parameter
$\zeta$ in \Figref{fig:5}. This figure in turn indicates a linear relation
between $K$ and $\zeta$ in the form $K\approx0.82\zeta$. In summary, we find
that the density of inertial impurities eventually becomes linearly correlated
with the vorticity, the regression coefficient being proportional to the
parameter $\zeta$.

\subsection{Turbulent mixing and clustering}
The behavior just described can easily be explained using an argument of
turbulent mixing, or turbulent equipartition.\cite{yank94,isic95,isic97}
Performing basic algebraic manipulations on \Eqref{eq:impur} while
accounting for the particular definition of the Lagrangian derivative $\DDt{}$
in \Eqref{eq:ddt}, we may rewrite the impurity model equation in the form
\begin{equation}\label{eq:impur2}
\DDt{(\ln\theta-\zeta\omega)}
=\zeta\nablab_\perp\ln\theta\cdotb\DDt{\nablab_\perp\varphi}+
\frac{\mu_\theta}{\theta}\,\nabla_\perp^2\theta,
\end{equation}
where the left-hand side describes advection by the electric drift and
compression by the polarization drift, while the right hand-side describes
advection by the polarization drift as well as collisional diffusion. We will
now argue that the right-hand side of this equation is small, which implies that
$\ln\theta-\zeta\omega$ is approximately a Lagrangian invariant, or in other
words, that $\ln\theta-\zeta\omega$ is nearly conserved along the trajectories
defined by the electric drift velocity. We split the impurity density $\theta$
into its mean value $\theta_0$ and the fluctuations $\theta_1$ arising from
compressibility. Based on the computational results, we further postulate
that $\theta_1/\theta_0=\order{\zeta}$. Selectively applying the split
$\theta\equiv\theta_0+\theta_1$ to \Eqref{eq:impur2}, we obtain
\begin{equation*}
\DDt{(\ln\theta-\zeta\omega)}
=\zeta\nablab_\perp\ln\biggl(1+\frac{\theta_1}{\theta_0}\biggr)\cdotb\DDt{\nablab_\perp\varphi}+
\frac{\mu_\theta}{\theta}\,\nabla_\perp^2\theta_1.
\end{equation*}
Assuming that the normalizations used in the impurity model adequately represent
the scales of fluctuations, the right-hand side of this last equation is of
order $\zeta^2+\zeta\mu_\theta$, whereas the fluctuating component of the
quantity inside the Lagrangian derivative on left-hand side is of order
$\zeta$. Hence, $\ln\theta-\zeta\omega$ varies slowly along the electric drift
trajectories, thus constituting an approximate Lagrangian invariant.

Turbulent mixing homogenizes Lagrangian invariants, and thus tends to drive
systems to states of so-called turbulent equipartition. The application of this
argument to the approximate invariant here yields
\begin{equation*}
\ln\theta-\zeta\omega\approx\mathrm{const}.
\end{equation*}
Because the amount of impurities is conserved and it is assumed that
$\theta_1/\theta_0=\order{\zeta}$, the previous expression can be recast into
the simpler form
\begin{equation*}
\theta/\theta_0\approx 1+\zeta\omega.
\end{equation*}
The present analysis thus predicts a linear relation between the impurity
density and the vorticity, the regression coefficient being the parameter
$\zeta$. Hence, the theoretical argument explains the effect discovered in the
simulations up to a slight mismatch in the regression coefficient. We emphasize
that the particular features of the HW turbulence model were not used in the
previous argument, the conclusions relying solely on the modeling of the
dynamics of impurities and the argument of turbulent mixing. It follows that the
clustering of inertial impurities in vortical structures is a generic effect,
independent of the underlying instability mechanisms and the specific
characteristics of the turbulence.

The fact that the density of inertial impurities becomes directly correlated
with the vorticity implies that impurities of positive charge cluster in
positive vortices, the opposite taking place for negatively charged impurities.
As previously introduced, this effect is similar to the aggregation of heavy
particles in anticyclonic vortices that can take place in rotating fluids as a
result of the Coriolis force.\cite{prov99} In contrast with compression
by the polarization drift, which yields $\DDt{\ln\theta}\sim\zeta\DDt{\omega}$,
compression by the Coriolis force enters the dynamics of heavy impurities in the
form $\DDt{\ln\theta}\sim-2\Omega\omega$, with $\Omega$ being the overall
rotation. Hence, aggregation in the rotating-fluid case does not necessarily
lead to a linear relation between impurity density and vorticity, but can
go on further to form highly concentrated clusters in the cores of
anticyclonic vortices. For this reason, it has for example been suggested as a
mechanism, in combination with gravitation, for planetary formation in
rotating astrophysical disks.\cite{brac99}

\section{Pinch of inertial impurities}\label{sec:pinch}
In this section we finally investigate the possible role of inertia in the
radially inward pinch that concentrates impurities in the center of the plasma
column in magnetic confinement devices.\cite{wagn93,dux03} As a measure of the
overall drift of impurities, we consider the global impurity flux
\begin{equation*}
\vec{\Gamma}_\theta \equiv \vec{\Gamma}_\theta^E + \vec{\Gamma}_\theta^p
\equiv\intd{\vec{x}}{\theta\vec{v}_E} +
\intd{\vec{x}}{\theta\vec{v}_p}.
\end{equation*}
Here, we have distinguished the components respectively resulting from advection
by the electric drift and by the polarization drift. In terms of these global
fluxes, we define the impurity drift velocity $\vec{V}_\theta$ and its
components $\vec{V}_\theta^E$ and $\vec{V}_\theta^p$ by means of
\begin{equation*}
\vec{V}_\theta\equiv\vec{V}_\theta^E+\vec{V}_\theta^p\equiv
Q^{-1}\vec{\Gamma}_\theta^E+Q^{-1}\vec{\Gamma}_\theta^p,
\end{equation*}
where once again $Q\equiv\intd{\vec{x}}\theta$. For a homogeneous distribution
of impurities in a periodic domain, both $\vec{\Gamma}_\theta^E$ and
$\vec{\Gamma}_\theta^p$ are zero. Hence, ideal impurities cannot experience a
net drift when their initial distribution is homogeneous, since such a 
distribution remains unchanged in the ideal case. In contrast, inertial
impurities may experience a drift even if their initial distribution is uniform,
given that the latter is altered by compressible effects, as shown in
the preceding section.

\subsection{Simulation results}
\begin{figure}
\includegraphics{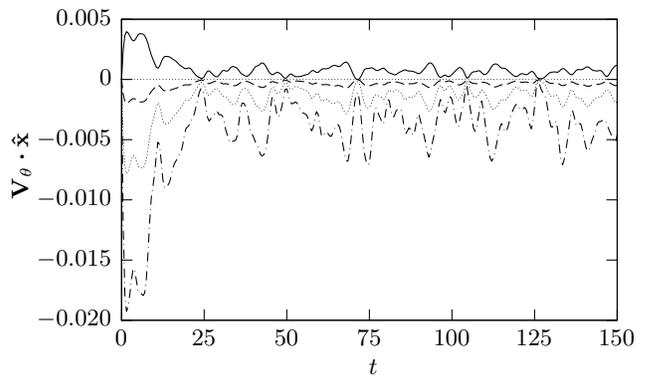}
\caption{\label{fig:6}%
Evolution of the radial drift velocity $\vec{V}_\theta\cdotb\uvec{x}$ of
inertial impurities in a HW saturated turbulent state with $\mathcal{C}=1$ 
for $\zeta=-0.01$ (solid), $\zeta=0.005$ (dashed), $\zeta=0.02$ (dotted), 
and $\zeta=0.05$ (dash-dotted).}
\end{figure}

\begin{figure}
\includegraphics{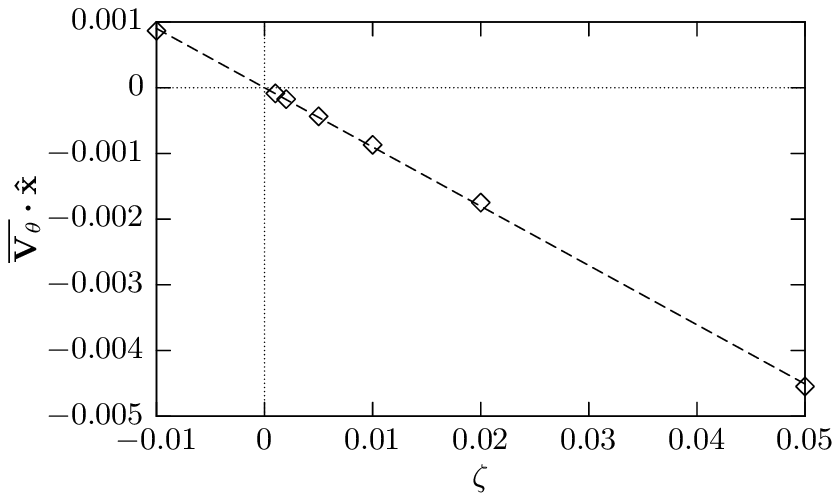}
\caption{\label{fig:7}%
Time-averaged radial drift velocity $\mean{\vec{V}_\theta}\cdotb\uvec{x}$ of inertial 
impurities as a function of $\zeta$ in a HW saturated turbulent with $\mathcal{C}=1$. 
The fitted line (dashed) corresponds to $\mean{\vec{V}_\theta}\cdotb\uvec{x}=-0.090\zeta$.}
\end{figure}

We begin by monitoring the radial drift velocity $\vec{V}_\theta\cdotb\uvec{x}$
in the simulations referred to in \Secref{sec:cluster}. These were initialized
with homogeneous distributions of inertial impurities in a HW saturated
turbulent state with $\mathcal{C}=1$.  In \Figref{fig:6} we present the
evolution of the radial drift velocity for four different values of $\zeta$. In
every case, the evolution of the radial drift comprises a strong transient
burst, after which the drift enters a saturated quasistationary
regime. Moreover, the radial drift velocity is seen to have a definite sign
opposite to that of $\zeta$, that is to say, opposite to the type of charge of
the impurity particles. It follows that positively charged impurities are
subject to a continuous radially inward drift, as observed in experimental
investigations of impurity transport.\cite{wagn93,dux03} In \Figref{fig:7} we
show the time-averaged radial drift velocities
$\mean{\vec{V}_\theta}\cdotb\uvec{x}$ in the saturated regimes, computed using
the values of the drifts between $t=25$ and $t=150$. The variation of the
time-averaged radial drift with $\zeta$ is remarkably linear, well fitted by
$\mean{\vec{V}_\theta}\cdotb\uvec{x}=-0.090\zeta$. Therefore, the present radial
drift scales linearly with the mass--charge ratio of the impurity particles.

The reason for this peculiar pinch effect is not clear at the outset. An
analysis of the components $\vec{V}_\theta^E$ and $\vec{V}_\theta^p$ of the
drift velocity in the simulations reveals that the contribution of the electric
drift is by far dominant. Hence, while the net drift arises as a result of the
polarization drift, the main role of the latter is not to globally advect
impurities, but rather to distribute them in such a way that they experience net
transport by the electric drift. From a naive point of view, it could be thought
that the correlation between impurity density and vorticity discovered in
\Secref{sec:cluster} is responsible for the radial drift. Indeed, transport of
trapped particles by coherent vortices is for instance known to play an
important role in transport processes in rotating fluids.\cite{prov99} However,
because an incompressible flow in a 2D periodic domain cannot globally transport
its vorticity, a purely linear relation between impurity density and vorticity
would yield no drift at all. Thus, the explanation for the radial drift of
inertial impurities requires deeper thought.

\subsection{Net transport and pinch}
Intrigued by the initial burst and the subsequent saturation of the radial
impurity drift velocity, we analyze the evolution of the global impurity
flux. In particular, we focus on the $\vec{\Gamma}_\theta^E$ component, given
that the net transport of impurities is dominated by the electric drift. In a
periodic domain, the evolution of $\vec{\Gamma}_\theta^E$ under the impurity
model in \Eqref{eq:impur} is governed by
\begin{equation*}
\ddt{\vec{\Gamma}_\theta^E} =\vec{\Lambda}_1+\vec{\Lambda}_2+\vec{\Lambda}_3+\vec{\Lambda}_4,
\end{equation*}
where
\begin{align*}
\vec{\Lambda}_1&\equiv
\intd{\vec{x}}{\theta\DDt{\vec{v}_E}},\\[0.5ex]
\vec{\Lambda}_2&\equiv
\intd{\vec{x}}{\zeta\theta\vec{v}_E\DDt{\omega}},\\[0.5ex]
\vec{\Lambda}_3&\equiv
\intd{\vec{x}}{\zeta(\nablab_\perp\theta\cdotb\DDt{\nablab_\perp\varphi})\vec{v}_E},\\[0.5ex]
\vec{\Lambda}_4&\equiv
\intd{\vec{x}}{\mu_\theta\vec{v}_E\nabla_\perp^2\theta}.
\end{align*}
These four contributions correspond respectively to evolution of the electric
drift velocity, compression by the polarization drift, advection by the
polarization drift, and collisional diffusion of impurities. In order to reveal
the magnitude of the different terms, we again split the impurity density
into its mean value $\theta_0$ and the fluctuations $\theta_1$, keeping in mind
that $\theta_1/\theta_0=\order{\zeta}$. Selectively applying the split
$\theta\equiv\theta_0+\theta_1$ to the above definitions we obtain
\begin{align*}
\vec{\Lambda}_1&=
\intd{\vec{x}}{\theta_1\DDt{\vec{v}_E}}=\order{\zeta},\\[0.5ex]
\vec{\Lambda}_2&=
\intd{\vec{x}}{\zeta\theta\vec{v}_E\DDt{\omega}}=\order{\zeta},\\[0.5ex]
\vec{\Lambda}_3&=
\intd{\vec{x}}{\zeta(\nablab_\perp\theta_1\cdotb\DDt{\nablab_\perp\varphi})\vec{v}_E}=\order{\zeta^2},\\[0.5ex]
\vec{\Lambda}_4&=
\intd{\vec{x}}{\mu_\theta\vec{v}_E\nabla_\perp^2\theta_1}=\order{\zeta \mu_\theta}.
\end{align*}
According to these estimates, $\vec{\Lambda}_1$ and $\vec{\Lambda}_2$ dominate
the dynamics of the global impurity flux. In the case of uniform impurity
density, $\vec{\Lambda}_2$ is the only nonvanishing contribution to
$\mathrm{d}\vec{\Gamma}_\theta^E/\mathrm{d}t$, being therefore responsible for
the initial development of drift velocities.

We will now resort to the particular HW turbulence model to show that
$\vec{\Lambda}_2$ constitutes a source for global radial flux in the present
case. Substitution of the HW vorticity equation in the above expression for
$\vec{\Lambda}_2$ yields
\begin{align*}
\vec{\Lambda}_2 &=
\intd{\vec{x}}{\zeta\theta\bigl[\mathcal{C}(\varphi-n)+\mu_\omega\nabla_\perp^2\omega\bigr]
\vec{v}_E}\\
&=-\zeta\mathcal{C}\theta_0\intd{\vec{x}}{n\vec{v}_E} + \order{\zeta^2+\zeta\mu_\omega}.
\end{align*}
Projecting this equation onto the radial direction we find
\begin{align*}
\vec{\Lambda}_2\cdotb\uvec{x}
&=-\zeta\mathcal{C}\theta_0\intd{\vec{x}}{n\pp{y}\varphi}+\order{\zeta^2+\zeta\mu_\omega}\\
&=-\zeta\mathcal{C}\theta_0\Gamma_n + \order{\zeta^2+\zeta\mu_\omega}.
\end{align*}
Here, we have identified the sole source $\Gamma_n$ of energy and generalized
enstrophy within the HW model. Since $\Gamma_n$ is in practice positive definite
(see, e.g., Table~I in \Ref{cama95}), it follows that
$\vec{\Lambda}_2\cdotb\uvec{x}$ has a definite sign opposite to that of $\zeta$,
thus acting as a source for the global radial flux of inertial impurities. As
explained before, $\vec{\Lambda}_2$ is initially the only nonzero contribution
to $\mathrm{d}\vec{\Gamma}_\theta^E/\mathrm{d}t$, thus explaining the transient
bursts observed in \Figref{fig:6}. More generally, this term sustains
the radial transport of impurities in the turbulent states described by the HW model.

From the above estimates, we are led to expect that $\vec{\Lambda}_1$ is
responsible for the saturation of the radial impurity drift seen in
\Figref{fig:6}. The reason for this saturation becomes clear when we
account for the correlation that develops between impurity density and
vorticity. The argument of turbulent equipartition used in \Secref{sec:cluster}
predicted an eventual relation $\theta_1/\theta_0\approx\zeta\omega$ between the
impurity density fluctuations and the vorticity. Considering this relation and
the fact that vorticity is not globally transported by the electric drift we
obtain
\begin{align*}
\vec{\Lambda}_1&\approx \intd{\vec{x}}{\zeta\theta_0\omega\DDt{\vec{v}_E}}\\
&=\zeta\theta_0\ddt{}\intd{\vec{x}}{\omega\vec{v}_E}-\zeta\theta_0\intd{\vec{x}}{\vec{v}_E\DDt{\omega}}\\
&=-\intd{\vec{x}}{\zeta\theta_0\vec{v}_E\DDt{\omega}}\approx-\vec{\Lambda}_2
\end{align*}
whenever turbulent equipartition holds true. Consequently, when such a state is
reached, the two terms dominating the evolution of the radial drift
approximately cancel, and the radial drift enters a quasistationary regime. The
time it takes for this regime to be reached is independent of $\zeta$, as the
development of correlation is exclusively dependent on the turbulent mixing by
the electric drift. In fact, the evolution of $\theta_1$ is approximately linear
with $\zeta$ within short time intervals, since in line with the discussion in
\Secref{sec:cluster} it follows that
\begin{equation*}
\DDt{\theta_1}=\zeta\theta_0\DDt{\omega}+\order{\zeta^2+\zeta\mu_\theta}.
\end{equation*}
Given that solely the density fluctuations yield net transport of impurities,
the evolution of the impurity drift is likewise approximately linear with
$\zeta$. This explains the apparent linear scaling of the curves in
\Figref{fig:6} and the linearity of the time-averaged radial drift in
\Figref{fig:7}.

While the arguments presented here do not yield a prediction for the
quasistationary level of the radial impurity drift, they do identify the
mechanism for its onset and likewise explain its scaling with the parameter
$\zeta$. We emphasize that the former mechanism relies on the specific
properties of the turbulence described by the HW model. Nevertheless, an
analogue pinch effect would appear for other driving instabilities provided that
the term $\vec{\Lambda}_2$ remains a source for a definite impurity drift.

\section{Conclusion}\label{sec:conclus}
We have studied the transport of impurity particles in plasma edge turbulence,
basing our investigations on the paradigmatic Hasegawa--Wakatani (HW) model for
resistive drift-wave turbulence and a consistent passive-fluid model for the
impurities. The latter model takes into account impurity-particle inertia, which
enters the impurity flow velocity in the form of the polarization drift. Inertia
has indeed been found to play a significant role in the turbulent transport of
impurities, since it gives rise to subtle yet qualitatively important
compressible effects.

The model equations have been investigated both theoretically and
computationally. The numerical simulations were performed using the
semi-Lagrangian (SL) pseudospectral code briefly described in the
Appendix. Although the emphasis of this paper is not on the numerical methods
employed, we point out the suitability of SL schemes for the advection-dominated
problems that appear in plasma turbulence. In particular, we highlight the
superior stability of the SL semi-implicit scheme used here for the HW model,
which is known to constitute a numerically challenging nonlinear problem.

In the case of ideal impurities, we have focused on their turbulent diffusion by
the electric drift. This effect has been demonstrated by relative-diffusion
analysis of the evolution of impurity puffs. The effective diffusivities of the
turbulent advecting field have been calculated for several values of the
adiabaticity parameter in the HW model. The resulting turbulent diffusivities
are in qualitative agreement with those obtained in an earlier study based on an
absolute-diffusion, test-particle approach.\cite{basu03b}

In the case of inertial impurities, we have discovered that their density
eventually correlates with the vorticity of the electric drift velocity. This
clustering effect scales linearly with the mass--charge ratio of the impurity
particles and results from compression by the polarization drift in combination
with turbulent mixing. The clustering of impurities in vortices of definite sign
is a generic effect in magnetized-plasma turbulence, in the sense that it is
independent of the specific characteristics of the turbulence.

Finally, we have also found an overall radial drift of inertial impurities that
scales linearly with the mass--charge ratio of the impurity particles and is
inward for positively charged impurities. This anomalous pinch relies on the
particular features of the turbulence described by the HW model. In contrast
with other known turbulent pinches,\cite{garb04} the present pinch is neither
caused by a specific magnetic field geometry nor by temperature gradients, given
that our models assume a uniform magnetic field and cold impurities. Consequently,
our finding of a net radial drift due to impurity-particle inertia constitutes
a new contribution towards the understanding of the anomalous pinch of
impurities in magnetic confinement devices.

\begin{acknowledgments}
The first author is grateful to M.~P.~S{\o}rensen for co-supervising his
M.Sc.~thesis, which constitutes the basis for this publication.
O.E.G. was sponsored by the Research Council of \mbox{Norway}.
\end{acknowledgments}

\appendix*
\section{Numerical schemes}
The numerical simulations presented in this paper were performed using a
semi-Lagrangian (SL) pseudospectral code developed by the first
author.\cite{prie04} In essence, SL schemes combine the very stable Lagrangian
treatment of advection with the convenience of a fixed spatial
discretization.\cite{stan91,boyd01} This is achieved by integrating the partial
differential equations in question along a different family of advective
trajectories in each time step, precisely those that by the end of the time step
traverse the fixed points on which the spatial discretization is based. We now
outline how the aforementioned code combines SL method and the pseudospectral
discretization to numerically solve the HW model and the impurity passive-fluid
model described in \Secref{sec:model}.

We discretize time using a constant time step $\Delta t$, which yields the
sequence $\{t_m\}$ of time instants related by $t_{m+1}=t_m+\Delta t$. Likewise,
we use the Fourier pseudospectral method to discretize the doubly periodic
rectangular simulation domain.\cite{boyd01} This results in a regular grid
$\{\vec{x}_i\}$ of collocation points. In our application of the SL method we
consider the (incompressible) electric drift velocity $\vec{v}_E=\uvec{z}\timesb\nablab_\perp\varphi$ as the sole
advecting field. Hence, the advective trajectory $\vec{X}(\vec{x}_i,t_{m+1};t)$
that passes through the grid point $\vec{x}_i$ at time $t_{m+1}$ is given by
\begin{align*}
\ddt{}\,&\vec{X}(\vec{x}_i,t_{m+1};t)=\vec{v}_E\bigl(\vec{X}(\vec{x}_i,t_{m+1};t),t\bigr),\\[1ex]
&\vec{X}(\vec{x}_i,t_{m+1};t_{m+1})=\vec{x}_i.
\end{align*}
At each time step, we firstly determine the so-called upstream points
$\{\vec{X}(\vec{x}_i,t_{m+1};t_{m})\equiv \vec{x}_i-\vec{\alpha}_i\}$. These
are numerically calculated using the following second-order scheme, which
combines the implicit midpoint rule and linear extrapolation of the advecting
field:\cite{temp87,mcdo87}
\begin{gather*}
\vec{\alpha}_i = \Delta t\vec{v}_E^*\bigl(\vec{x}_i - \tfrac{1}{2}\vec{\alpha}_i, t_{m+1/2}\bigr),\\[1ex]
\vec{v}_E^*(\vec{x}, t_{m+1/2}) = \tfrac{3}{2}\vec{v}_E(\vec{x}, t_m) -
\tfrac{1}{2}\vec{v}_E(\vec{x}, t_{m-1}),
\end{gather*}
where $t_{m+1/2}=t_m+\tfrac{1}{2}\Delta t$. The implicit equations for the
displacements $\{\vec{\alpha}_i\}$ are solved using Newton's method, which 
converges in very few iterations.

Subsequently, the model equations are integrated from $t_m$ to $t_{m+1}$ along
the advective trajectories. For the HW model, \Eqref{eq:hw}, we use the
following unconditionally stable second-order semi-implicit scheme:
\begin{align*}
n^{m+1}-\widetilde{n}^{m} &=\tfrac{1}{2}\Delta t\mathcal{C}\Bigl[(\varphi-n)^{m+1}+ \widetilde{(\varphi-n)}{}^{m}\Bigr]\\
&\quad+\tfrac{1}{2}\Delta t\mu_n\Bigl[\nabla_\perp^2 n^{m+1}+\widetilde{\nabla_\perp^2 n}{}^{m}\Bigr]\\
&\quad+x^{m+1}-\widetilde{x}^{m},\displaybreak[0]\\[1ex]
\omega^{m+1}-\widetilde{\omega}^{m}
&= \tfrac{1}{2}\Delta t \mathcal{C}\Bigl[(\varphi-n)^{m+1}+\widetilde{(\varphi-n)}{}^{m}\Bigr]\\
&\quad+\tfrac{1}{2}\Delta t \mu_\omega\Bigl[\nabla_\perp^2\omega^{m+1}+\widetilde{\nabla_\perp^2 \omega}{}^{m}\Bigr].
\end{align*}
Here the superscripts indicate time, and tildes denote evaluation at the
upstream points. Hence, for a function $\vec{g}(\vec{x},t)$ we have
$\widetilde{\vec{g}}{}^m(\vec{x}_i)=\vec{g}(\vec{x}_i-\vec{\alpha}_i,t_m)$. The
above equations supplemented with $\omega^{m+1}=\nabla_\perp^2\varphi^{m+1}$ are
easily solved within the Fourier discretization.

For the impurity passive-fluid model, \Eqref{eq:impur}, we employ the following
second-order implicit-explicit scheme:
\begin{gather*}
\theta^{*}-\widetilde{\theta}^{m} = 
-\Delta t\widetilde{\nablab_\perp\cdotb(\theta\vec{v}_p)}{}^{m}
+\tfrac{1}{2}\Delta t \mu_\theta \Bigl[\nabla_\perp^2 \theta^{*}
+\widetilde{\nabla_\perp^2 \theta}{}^{m}\Bigr],\displaybreak[0]\\[1ex]
\begin{split}
\theta^{m+1}-\widetilde{\theta}^{m} &= 
-\tfrac{1}{2}\Delta t\Bigl[\nablab_\perp\cdotb(\theta^*\vec{v}_p)^{m+1}
+\widetilde{\nablab_\perp\cdotb(\theta\vec{v}_p)}{}^{m}\Big]\\
&\quad+\tfrac{1}{2}\Delta t \mu_\theta\Bigl[\nabla_\perp^2 \theta^{m+1} + \widetilde{\nabla_\perp^2 \theta}{}^{m}\Bigr].
\end{split}
\end{gather*}
In our code we further approximate the polarization drift
$\vec{v}_p=-\zeta\DDt{\nablab_\perp\varphi}$ by means of
\begin{equation*}
\vec{v}_p^{m+1}=\zeta\bigl(\widetilde{\nablab_\perp\varphi}{}^{m}-\nablab_\perp\varphi^{m+1}\bigr)/\Delta t+\order{\Delta t}.
\end{equation*}
While in theory this approximation reduces the accuracy of the scheme to first
order, its effect in practice is very limited given the smallness of the polarization drift.

We lastly note that the combination of spectral discretizations with SL schemes
is problematic.\cite{boyd01} The difficulty lies in the required evaluations at
midpoints and upstream points, which are generally not equispaced. Performing
such evaluations exactly, using the spectral representation, is in most cases
computationally infeasible. In our code this problem is circumvented using
high-order periodic spline interpolation on a refined grid.\cite{beyl95,stei98}


\end{document}